\documentclass[aps,prb,showpacs,twocolumn,citeautoscript,superscriptaddress]{revtex4}
\usepackage{graphicx}
\usepackage{bm}
\newcommand{\TN}{$T_{N}$}

\newcommand{ \cro}{\mbox{Cr$_2$O$_{3}$}}

\begin{document}

\title{Microscopic theory of temperature-dependent magnetoelectric effect in \cro}
\author{Maxim~Mostovoy}
\affiliation{Zernike Institute for Advanced Materials,
University of Groningen, Nijenborgh 4, The Netherlands}
\author{Andrea~Scaramucci}
\affiliation{Zernike Institute for Advanced Materials,
University of Groningen, Nijenborgh 4, The Netherlands}
\author{Kris~T.~Delaney}
\affiliation{Materials Department, University of California,
Santa Barbara, California 93106-5050, USA}
\author{Nicola~A.~Spaldin}
\affiliation{Materials Department, University of California,
Santa Barbara, California 93106-5050, USA}
\date{\today}

\begin{abstract}
We calculate the temperature-dependent magnetoelectric response of \cro\ from 
first principles. 
The form of the dominant magnetoelectric coupling is determined using symmetry 
arguments, its strength is found using {\em ab initio} methods, and the 
temperature dependence of the response is obtained from Monte Carlo simulations.
The quantitative agreement of our results with experiment shows that the strong 
temperature dependence of the magnetoelectric effect in \cro\ results from 
non-relativistic exchange interactions and spin fluctuations.
\end{abstract}

\maketitle

\noindent{\it Introduction}:
Recent progress in understanding the sub-class of multiferroics in which improper 
ferroelectric
polarizations are induced by non-centrosymmetric magnetic orderings has led to 
a clarification of the microscopic origins for magnetoelectric coupling
\cite{CheongNatMat2007}.
In particular two distinct coupling mechanisms have been identified.
The first arises from relativistic effects linking electron spin and orbital 
momentum, resulting in the antisymmetric $\mathbf{S}_1 \times \mathbf{S}_2$ 
interaction between spins of magnetic ions. 
The dependence of the strength of this Dzyaloshinksii-Moriya interaction on polar displacements of ions makes magnets with non-collinear spiral orders ferroelectric\cite{KatsuraPRL2005,SergienkoPRB2006,MostovoyPRL2006,MalashevichPRL2008}.
In the second mechanism, polar deformations of the lattice are induced by 
Heisenberg spin 
exchange interactions $\mathbf{S}_1 \cdot \mathbf{S}_2$, originating from the 
Fermi statistics of electrons\cite{SergienkoPRL2006,PicozziPRL99}. 
This non-relativistic mechanism can give rise to stronger magnetoelectric couplings 
than those resulting from relativistic effects, which tend to be relatively weak in 
$3d$ transition metal compounds, and is not restricted to non-collinear spin 
arrangements.
Indeed, the electric polarizations of the orthorombic 
manganite Y$_{1-x}$Lu$_x$MnO$_3$\cite{Ishiwata2009} and orthoferrite 
GdFeO$_3$\cite{TokunagaNatMat2009}, which have collinear antiferromagnetic spin 
orderings and polarization arising from the Heisenberg mechanism, exceed the
largest polarizations observed in spiral multiferroics by one order of 
magnitude.

In this work we show that, in addition to causing  multiferroic behavior, the relativistic 
and Heisenberg exchange mechanisms can both give rise to the linear magnetoelectric effect, 
in which an applied magnetic field induces an electric polarization proportional to 
the field and, conversely, magnetization is induced by an applied electric field 
(for a review see Ref. \onlinecite{FiebigJAPD2005}). Furthermore, the two mechanisms 
can co-exist in the same material and dominate in different temperature regimes. 
Using a combination of first-principles density functional theory 
and Monte Carlo methods we calculate the temperature-dependent
magnetoelectric response of the prototype magnetoelectric material, 
chromium sesquioxide, \cro.
We show that the strong finite-temperature magnetoelectric response
originates from the Heisenberg exchange mechanism combined with thermal 
spin fluctuations, whereas at zero temperature, where spin fluctuations 
vanish, the observed weak response arises from relativistic effects.
Significantly, \cro\ is a collinear antiferromagnet, and so our work extends
the recent suggestion that Heisenberg exchange interactions can give rise to a 
strong magnetoelectric effect in frustrated magnets in which 
spins are forced to form non-collinear orders with nonzero toroidal or
monopole moments\cite{DelaneyPRL2009,SpaldinJPCM2008}. 

\noindent{\it Magnetoelectric coupling}:
\begin{figure}[tbp]
\centering
\includegraphics[width=0.35\textwidth]{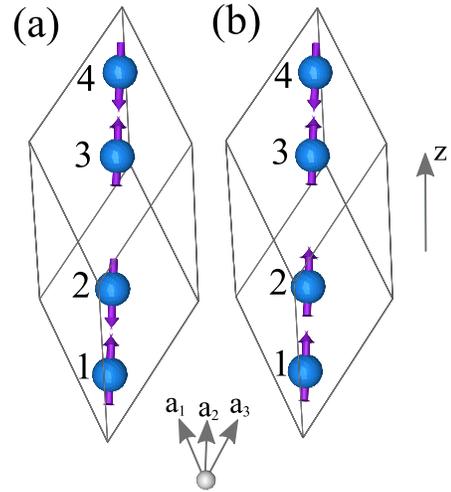}
\caption{(Color online) Rhombohedral unit cell of \cro\ with the unit vectors $\mathbf{a}_i, i = 1,2,3$, containing four magnetic Cr ions. Panel (a) shows the actual antiferromagnetic spin ordering in \cro, while panel (b) shows the spin ordering imposed in our first-principle calculations to induce an electric polarization along the trigonal $z$ axis.}
\label{fig:unitcell}
\end{figure}
The magnetoelectric effect in \cro\ was predicted phenomenologically by 
Dzyaloshinksii\cite{DzyaloshinskiiJETP1959} and measured by 
Astrov\cite{AstrovJETP1960} shortly after the theoretical prediction. 
\cro\  has  four Cr$^{3+}$ ions with spin $S = 3/2$ in the rhombohedral unit cell. Below \TN\ = 307K it shows collinear $\uparrow\downarrow\uparrow\downarrow$ spin ordering along the trigonal $z$ axis\cite{BrockhouseJChemPhys}  [see Fig.~\ref{fig:unitcell} (a)]. This antiferromagnetic ordering preserves the three-fold and two-fold symmetry axes of the paramagnetic phase, but it breaks inversion symmetry $I$, reducing
it to $I'$ (inversion combined with time reversal), which allows for two independent magnetoelectric coupling terms in the free energy\cite{DzyaloshinskiiJETP1959}:
\begin{equation}\label{eq:freeenergy}
   F_{\rm me} = -g_{\parallel} G^z E^z H^z
    - g_{\perp} G^z \left(E^x H^x + E^y H^y \right),
\end{equation}
where $G^z = \left \langle S_{1}^z - S_{2}^z + S_{3}^z - S_{4}^z \right\rangle$ is the antiferromagnetic order parameter. The two magnetoelectric coefficients, $\alpha_{\parallel} = g_{\parallel} G^z$ and $\alpha_{\perp} = g_{\perp} G^z$, show very different temperature dependences\cite{AstrovJETP1960}. While $\alpha_{\perp}$ varies only in the immediate vicinity of the transition temperature, $T_N$, the coefficient $\alpha_{\parallel}$ is very strongly temperature dependent. $\alpha_\parallel$ reaches its maximum at $T_{\rm max}\sim260$K, below which it steeply decreases and changes sign at $\sim100$K. While at $T_{\rm max}$ the magnitude of $\alpha_{\parallel}$ is one order of magnitude larger than $\left|\alpha_{\perp}\right|$, at low temperatures
$\left|\alpha_{\parallel}\right| < \left|\alpha_{\perp}\right|$. 

Early measurements of the temperature dependence  of the magnetoelectric coefficients\cite{RadoPR1962,HornreichPRB1967,YatomPR1969}, as well as recent first-principles calculations\cite{IniguezPRL2008}, show that the relatively weak magnetoelectric effects at low temperature 
result from relativistic interactions. In what follows, we demonstrate that the much stronger response at elevated temperatures originates from Heisenberg exchange.

Phenomenologically, the coupling of the electric polarization $P^z$ along the trigonal axis to spins, resulting from the dependence of spin-exchange constants on polar lattice distortions, is described by
\begin{equation}\label{eq:polarization}
    P^z = \lambda \left ( \mathbf{S}_{1} \cdot \mathbf{S}_{3} -
    \mathbf{S}_{2} \cdot \mathbf{S}_{4}\right),
\end{equation}
where $\mathbf{S}_{\alpha}$ with $\alpha=1,2,3,4$ denotes the sublattice magnetization and $\lambda$ is the coupling strength that we will determine from \textit{ab initio} calculations. Here we take into account the fact that the exchange-driven polarization can only depend on scalar products of the magnetizations. The combination of scalar products in the right-hand side of  Eq.(\ref{eq:polarization}) transforms as $P^z$, which changes sign under $C_2$ and $I$. The combination of scalar products
that lead to this property can be seen by inspection of Table~\ref{tab:sitetransformation}, which shows how the four inequivalent magnetic sites transform under the symmetry operations of \cro.
\begin{table}[htbp]
\centering
\begin{tabular}{|c|c|c|c|}
\hline
  & $C_3$ & $C_2$ &  $I$\\
[0.4ex] \hline
\,\,\,\,\,\,1\,\,\,\,\,\,&\,\,\,\,\,\,1\,\,\,\,\,\,& 2  & \,\,\,\,\,\,4\,\,\,\,\,\,\\
2& 2 & 1 & 3\\
3& 3 & $\,4-c\,$ & 2\\
4& 4 & $3-c$ & 1\\
[0.1ex] \hline
\end{tabular}
\caption{Transformation of four independent Cr sites with the fractional coordinates $\mathbf{r}_1 = (u,u,u)$, $\mathbf{r}_2 = (1/2-u,1/2-u,1/2-u)$, $\mathbf{r}_3 = (1/2+u,1/2+u,1/2+u)$, and $\mathbf{r}_4 =  (1-u,1-u,1-u)$, where $u \approx 0.153$, under the generators of space group $R{\bar3}c$: the $120^{\circ}$-rotation around the $z$ axis, $C_3 = (x_3,x_1,x_2)$, the $180^{\circ}$-rotation around the axis orthogonal to the $z$-direction, $C_2 = (1/2-x_2,1/2-x_1,1/2-x_3)$ and inversion $I = (1-x_1,1-x_2,1-x_3)$. Here, $\mathbf{c} = \mathbf{a}_1 + \mathbf{a}_2 + \mathbf{a}_3$, where $\mathbf{a}_i$ ($i=1,2,3$) are the rhombohedral unit vectors.}
\label{tab:sitetransformation}
\end{table}

Equation (\ref{eq:polarization}) is clearly appropriate for describing electric polarization induced by spin ordering in a multiferroic material. In addition, it applies to the linear magnetoelectric effect, as can be seen from the following heuristic argument: In an applied magnetic field $H^z$ the average value of spin on the sublattice $\alpha$ changes by $\left \langle \delta S_{\alpha}^{z} \right \rangle \propto \chi_{\parallel} H^z$, where $\chi_{\parallel}$ is the longitudinal magnetic susceptibility. Equation (\ref{eq:polarization}) then gives $P^z \propto \lambda \chi_{\parallel} \left\langle S_1^z - S_2^z + S_3^z - S_4^z \right\rangle H^z \propto G^z H^z$, consistent with $P^z = - \frac{\partial F_{\rm me}}{\partial E^z} = g_{\parallel} G^z H^z$ obtained from Eq.(\ref{eq:freeenergy}).

Equation (2) is only meaningful within the mean field approach. To account for effects of spin fluctuations on the magnetoelectric response of \cro\ we will use the microscopic expression for the exchange-driven polarization in terms of scalar products of Cr spins (rather than the sublattice magnetizations), which has the form
\begin{equation}
    P^z =  \frac{\lambda}{6N}   \sum_{j} \sum_{n=1}^{6}
    \left(
    \mathbf{S}_{1,j} \cdot \mathbf{S}_{3,j-b_{n}}-
    \mathbf{S}_{2,j} \cdot \mathbf{S}_{4,j-b_{n}}
    \right).
    \label{eq:fullcoupling}
\end{equation}
Here $j$ labels unit cells, $N$ is the total number of unit cells, and $\mathbf{b}_1 = \mathbf{a}_1$, $\mathbf{b}_2 = \mathbf{a}_2$, $\mathbf{b}_3 = \mathbf{a}_3$, $\mathbf{b}_4 = \mathbf{a}_1+\mathbf{a}_2$,
$\mathbf{b}_5 = \mathbf{a}_2+\mathbf{a}_3$ $\mathbf{b}_6 = \mathbf{a}_3+\mathbf{a}_1$ ($\mathbf{a}_i$ being the rhombohedral unit vectors).  Remarkably, the interaction between the fourth nearest-neighbor Cr ions separated by the distance  $r_{1,(3 - a_1)} = r_{1,(3 - a_1 - a_2)} = 3.65$\AA\  turns out to give rise to the magnetoelectric effect in \cro. The shorter-range exchange interactions do not couple the sublattices 1 and 3 (or 2 and 4) and, therefore, do not contribute to $P^z$, while other interactions between these sublattices correspond to much longer exchange paths and are negligibly small. This allows us to accurately pinpoint the microscopic origin of the strong magnetoelectric effect in \cro.

Importantly, Eqs.(\ref{eq:polarization}) and (\ref{eq:fullcoupling}) apply to any four-sublattice spin ordering in \cro. The $\uparrow\downarrow\uparrow\downarrow$ spin ordering realized in the
low-temperature ground-state of \cro\ induces no electric polarization, and so a straightforward density-functional study of \cro\ does not give information about the strength of the finite-temperature perpendicular magnetoelectric coupling. However, the $\uparrow\uparrow\uparrow\downarrow$ ordering shown in Fig.~\ref{fig:unitcell} (b) renders \cro\ multiferroic and induces the electric polarization
\begin{equation}\label{eq:Pcalc}
    \mathcal{P} \equiv P^z(\uparrow\uparrow\uparrow\downarrow) = 2 \lambda S^2.
\end{equation}
We next extract $\lambda$ using {\em ab initio} methods by enforcing $\uparrow\uparrow\uparrow\downarrow$ spin ordering and calculating the magnetically-induced polarization $\mathcal{P}$.

\noindent{\it First-principles calculations of the magnetoelectric coupling}:
We compute $\lambda$ and the spin exchange parameters using plane-wave density-functional theory, as implemented in the
Vienna Ab-initio Simulation Package (\verb!VASP!)\cite{VASP}. We use PAW potentials\cite{PAW} for core-valence
partitioning, and the local-spin-density approximation with a rotationally invariant 
Hubbard-$U$ (LSDA$+U$) for the exchange-correlation 
potential\cite{SashaLDAU}. Our Hubbard $U=2.0$\,eV is the same value that was taken for computing the magnetoelectric response at
zero temperature\cite{IniguezPRL2008}. We deliberately do not include spin-orbit coupling, and all
calculations are for collinear spin densities, so that $\lambda$ corresponds only
to polarizations induced by exchange striction.

We find that a plane-wave cutoff of $500$\,eV and Monkhorst-Pack\cite{MP} $k$-point sampling 
of $4\times4\times4$ are sufficient for computing the properties of interest.
Note that much finer $k$ meshes would be required for accurately resolving the magnetocrystalline 
easy axis\cite{IniguezPRL2008}, but non-relativistic properties are well converged with 
our chosen parameters.

We work with space group $R\bar{3}c$ at the experimental volume\cite{ExpStruct} of 96.0\AA$^3$ and rhombohedral angle of 55.13$^\circ$. The internal coordinates are relaxed within our 
density functional calculations for the $\uparrow\downarrow\uparrow\downarrow$ magnetic configuration, yielding coordinates $x=0.1536$ for Cr, and $x=0.9426$ for O in Wyckoff positions $4c$ and $6e$ 
respectively. Subsequently, the Heisenberg exchange couplings $J_1$---$J_5$\cite{SamuelsenPhysica1970},  corresponding to Cr-Cr distances of $2.65$ to $4.10$\,\AA{}, are computed by fitting a Heisenberg 
Hamiltonian to DFT total energies of twelve different spin configurations with fixed ion coordinates in the hexagonal setting of $R\bar{3}c$. This method is analogous to that employed by Shi {\it et al.}\cite{Shi}. 

Finally, we compute $\lambda$ by enforcing the spin configuration of $\uparrow\uparrow\uparrow\downarrow$ and re-relaxing
the ionic coordinates in the rhombohedral unit cell. 
The resulting ionic configuration has a polar lattice distortion. 
We compute the magnitude of $\mathcal{P} = 0.585$\,$\mu$C/cm$^2$ using the Berry phase approach\cite{Berry}, which allows us to extract $\lambda$. We note that $\mathcal{P}$ is of the same order of magnitude as the polarization induced by exchange interactions in multiferroics with collinear spins\cite{Ishiwata2009,TokunagaNatMat2009}.

\noindent{\it Monte Carlo simulations}:
Using Eq.(\ref{eq:fullcoupling}) we can express the magnetoelectric coefficient $\alpha_{\parallel}$ in terms of spin correlation functions:
\begin{equation}
 \alpha_{\parallel} =  \left. \frac{\partial \langle P^z \rangle}{\partial H^z}\right|_{H^z=0}= \frac{2 \mu_B}{k_B T}  \langle P^z \sum_{\alpha,j} S^z_{\alpha,j} \rangle,
\label{eq:alphaStat}
\end{equation}
where $\langle \ldots \rangle$ denotes the thermal average at temperature $T$, $k_B$ is the Boltzmann constant and $\mu_B$ is the Bohr magneton.

In the mean-field approximation, equivalent to replacing the scalar products of spins $\mathbf{S}_{\alpha,j} \cdot \mathbf{S}_{\beta,k}$ by $\langle \mathbf{S}_{\alpha,j} \rangle \cdot \mathbf{S}_{\beta,k} +\mathbf{S}_{\alpha,j} \cdot \langle\mathbf{S}_{\beta,k}\rangle$ in the expression for $P^z$, one obtains\cite{RadoPR1962,HornreichPRB1967}
\begin{equation}\label{eq:alphaMF}
    \alpha_{\parallel} = \frac{\lambda v_0 G^z
    \chi_{\parallel}}{8 \mu_{\rm B}},
\end{equation}
where $v_0$ is the unit cell volume, in agreement with the simple argument given above. The mean-field expression 
qualitatively explains the observed temperature dependence of $\alpha_{\parallel}$: It first grows, together with the order parameter $G^z$, as the temperature drops below $T_N$, then subsequently decreases and vanishes at $T=0$, together with the 
longitudinal magnetic susceptibility $\chi_{\parallel}$. 

In Ref. \onlinecite{YatomPR1969} an attempt was made to take into account the effects of spin fluctuations using a higher-order decoupling scheme. This approximation fails, however, close to the transition temperature where spin fluctuations are large. We include spin fluctuations by calculating $\alpha_{\parallel}$ numerically using Monte Carlo simulations of a system of $864$ classical spins with exchange constants and the magnetoelectric coupling $\lambda$ obtained from our first principles calculations, as described above.

Since the correlation function in the right-hand side of Eq.(\ref{eq:alphaStat}) is zero unless a single antiferromagnetic domain is selected, we apply to our finite  spin system a weak staggered field along the $z$ axis,  $h(-)^{\alpha}$, where $\alpha = 1,2,3,4$ labels magnetic sublattices. This mimics the easy magnetic anisotropy of \cro\ as well as the cooling in electric and magnetic fields, used in measurements of magnetoelectric coefficients to select the domain with a given sign of magnetic order parameter. The field strength, $h=0.165$\,meV, was chosen so that it is small compared to the scale of exchange interactions, but large enough to make the Monte Carlo results independent of $h$.

\begin{figure}[Htbp]
\centering
\includegraphics[width=0.50\textwidth]{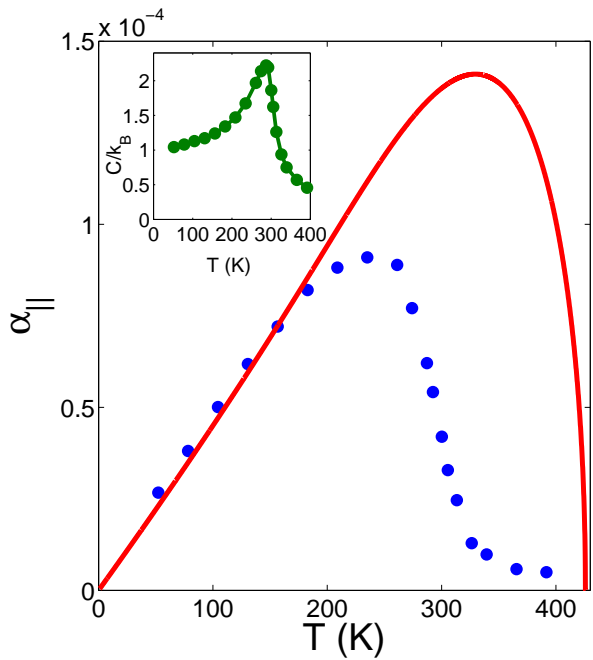}
\caption{(Color online) Temperature dependence of the magnetoelectric coupling $\alpha_{\parallel}$ obtained using {\em ab initio} values of the exchange constants and magnetoelectric coupling combined with Monte Carlo simulations (blue circles) and mean field calculations (solid red line). The inset shows the temperature dependence of magnetic specific heat.}
\label{fig:alphaT}
\end{figure}

Figure~\ref{fig:alphaT} shows the temperature dependence of $\alpha_{\parallel}$ obtained from Monte Carlo simulations (blue circles) and in the mean-field approximation described above (red solid line). The onset of the magnetoelectric response in our Monte
Carlo simulations, as well as the sharp peak in the specific heat (inset), shows that the antiferromagnetic order sets in at $\sim 290$\,K, close to the experimentally observed transition temperature $T_N=307$\,K. The maximal value of the 
magnetoelectric coefficient obtained from our simulations is $0.9 \times 10^{-4}$ (in CGS units), in excellent agreement with the experimental value of $1.0 \times 10^{-4}$ (see Ref. \onlinecite{Borovik2003}). The maximum value is reached at $\sim 240$\,K that compares well to $T_{\rm max} \sim 260$\,K found in experiment. The mean field transition temperature ($425$\,K) and maximal $\alpha_{\parallel}$ are significantly higher than the Monte Carlo values, indicating the importance of spin fluctuations in this material.

\noindent{\it Conclusions}:
We have presented the first \textit{ab initio} calculations of temperature-dependent linear magnetoelectric responses. The quantitative agreement of our results with experimental data on Cr$_2$O$_3$ demonstrates that the dominant parallel 
magnetoelectric coupling in this material originates from non-relativistic exchange interactions between electrons. The strong temperature dependence of the magnetoelectric coefficient $\alpha_{\parallel}$ underscores the general importance of spin fluctuations for magnetoelectric responses of materials with collinear spin orders. The magnetic-field-induced electric polarization of \cro\ is comparable to that predicted recently for an exchange-interaction-driven Kagome antiferromagnet with non-collinear spin ordering \cite{DelaneyPRL2009}. However, the magnetoelectric response of non-collinear magnets does not vanish at zero temperature as it involves transverse rather than longitudinal magnetic susceptibility and originates from the dynamic `electromagnon' modes, which can be excited by both electric and magnetic fields and which are absent in collinear magnets \cite{VegteMostovoy,ValdesPRL2009}. 
 
The approach used in this paper, specifically the combination of first principles calculations for artificially imposed magnetic states to extract parameters with Monte Carlo simulations of physically interesting quantities, opens a route to theoretical studies of a large variety of temperature-dependent static and dynamic magnetoelectric phenomena. Accurate predictions of the magnitude of magnetoelectric responses at finite temperature will greatly facilitate the search for and design of materials with the strongest responses.

\begin{acknowledgments}
The work of MM and AS was supported by the Thrust II program of the Zernike Institute 
for Advanced Materials and by the Stichting voor Fundamenteel Onderzoek der Materie (FOM). 
KTD and NAS were supported by the National Science Foundation under 
Award No.~DMR-0940420.
This work made use of the computing facilities of the California Nanosystems Institute
with facilities provided by NSF grant No. CHE-0321368 and Hewlett-Packard.
\end{acknowledgments}

\end{document}